\documentclass[aps,pre,showpacs,preprintnumbers,amsmath,amssymb,twocolumn,superscriptaddress]{revtex4-1}

\pdfoutput=1

\usepackage{amsmath}
\usepackage{amssymb}
\usepackage{graphicx}
\usepackage{dcolumn}
\usepackage{tabularx}
\usepackage{soul}
\usepackage{color}
\usepackage{bm}
\usepackage{latexsym}

\usepackage{multirow}
\usepackage{hyperref}
\usepackage[normalem]{ulem}

\newcommand{\ve}[1][K]{\mathbf{#1}}

\begin{document}

\title{Local orientational mobility in regular hyperbranched polymers}
\date{\today}

\author{Maxim Dolgushev}
\email{dolgushev@physik.uni-freiburg.de}
\affiliation{Institute of Physics, University of Freiburg, Hermann-Herder-Str. 3, 79104 Freiburg, Germany}
\affiliation {Institut Charles Sadron, Universit\'e de Strasbourg and CNRS, 23 rue du Loess, 67034 Strasbourg Cedex, France}
\author{Denis A. Markelov}
\affiliation{St. Petersburg State University, 7/9 Universitetskaya nab., St. Petersburg, 199034, Russia} 
\affiliation{St. Petersburg National Research University of Information Technologies,
Mechanics and Optics (ITMO University), Kronverkskiy pr. 49, St. Petersburg,
197101, Russia}
\author{Florian F\"urstenberg}
\affiliation{Institute of Physics, University of Freiburg, Hermann-Herder-Str. 3, 79104 Freiburg, Germany}
\author{Thomas Gu\'erin}
\affiliation{Laboratoire Ondes et Mati\`ere d'Aquitaine (LOMA), CNRS UMR 5798, Talence, France}

\begin{abstract}
We study the dynamics of local bond orientation in regular hyperbranched polymers modeled by Vicsek fractals. The local dynamics is investigated through the temporal autocorrelation functions of single bonds and the corresponding relaxation forms of the complex dielectric susceptibility. We show that the dynamic behavior of single segments depends on their remoteness from the periphery rather than on the size of the whole macromolecule. Remarkably, the dynamics of the core segments (which are most remote from the periphery) shows a scaling behavior which differs from the dynamics obtained after structural average. We analyze the most relevant processes of single segment motion and provide an analytic approximation for the corresponding relaxation times. 
Furthermore, we describe an iterative method to calculate the orientational dynamics in the case of very large macromolecular sizes.
\end{abstract}

\maketitle

\section{Introduction}

Hyperbranched polymers (HP) are macromolecules with a large number of branching units \cite{yan11,gao04,voit09,segawa13}. In contrast to dendrimers, which are typically synthesized in multistep schemes, HP are created in single-step reactions \cite{yan11,gao04,voit09}, making them very attractive for applications \cite{yan11,gao04}.
From the theoretical point of view, {the HP built from a
reaction that allows cluster-cluster aggregation show scaling, whereas
the ones created using a procedure where monomers are added
sequentially to an existing core which strictly avoids cluster-cluster
aggregation (e.g., dendrimers) do not scale \cite{jurjiu14}.} We note that HP typically possess a high degree of structural polydispersity; this, however, does not {break} the feature of {possible} scaling \cite{sokolov02}. Therefore, deterministic fractal structures are a very useful tool to understand the properties of HP in depth  \cite{sokolov02,gurtovenko05,alexander82,rammal83,cates84,muthukumar85,cosenza92,jayanthi92,schiessel98,blumen03,blumen04,agliari08,zhang09,lin10,meyer12,reuveni08,reuveni10,reuveni12,reuveni12pre,polinska14,sokolov16,dolgushev15}.

The theory of HP dynamics has been intensively developed with the focus on mechanical relaxation, microrheology, and macroscopic dielectric relaxation \cite{sokolov02,gurtovenko05,schiessel98,blumen03,blumen04}. All these dynamic properties have a striking feature in common: in the generalized Gaussian scheme (GGS), they can be calculated based only on the eigenvalue spectrum of the connectivity matrix (which describes the topology of the links between the monomers). The reason for this is that they typically represent macroscopic or structurally averaged properties. 

However, {the dynamics of a monomer of a macromolecule is generally complex and influences its kinetics of binding with other reactants \cite{degennes82,wilemski74a,guerin12,dolgushev15}; in the case of HP the local monomer dynamics could then be relevant for applications such as drug delivery \cite{liu10} or catalysis \cite{schlenk00} (since the related processes of sorption and desorption would depend on the mobility of both the sorbed substance and the macromolecule's monomers). Experimentally, the local dynamics can be studied by considering the local bond dynamics in NMR or dielectric relaxation experiments \cite{mijovic07,pinto13a,pinto13b,hofmann15,mohamed15}}. But the related local dynamical functions then depend on the particular location of the segments in the macromolecular structure. Consequently, the eigenvectors of the connectivity matrix (and not only the eigenvalues) influence the local dynamics, a fact that considerably complicates the computations of local dynamic functions in the case of large macromolecular sizes. To our knowledge, temporal autocorrelation functions of the spatial distance between monomers (after disorder averaging) have been investigated in the context of proteins  in Refs. \cite{granek05,reuveni12,reuveni12pre}, where they were shown to vary strongly with the chemical distance between the considered monomers. However, the dynamics of single segments, with focus on their location in the structure, has not been studied yet. Here we look at very local scale represented through single bonds and, in particular, on the influence of the bonds' location on their dynamics.

In this paper, we investigate the local dynamics of single bonds in HP modeled through Vicsek fractals {(VF)}, focusing on the local dielectric relaxation (for type A polymers in Stockmayer's classification \cite{stockmayer67},\footnote{{ The theoretical framework used here involves only longitudinal modes, and can thus be used to study dielectric relaxation of type A polymers, for which the dipoles are aligned with the bonds. Inclusion of rigid rodlike elements and associated with them splitting on longitudinal and transversal modes \cite{toshchevikov06} will allow one to study dielectric relaxations of other types (i.e., B and C).}}). We show that the dynamics of single segments strongly depends on their location in the macromolecule. Remarkably, the imaginary part of the local dielectric susceptibility of the core segments shows a different scaling than the global (structure-averaged) dielectric susceptibility. Indeed, while the dielectric susceptibility scales  as $\omega^{d_s/2}$ for the overall structure \cite{blumen04}
(with $d_s$ the spectral dimension of the structure and $\omega$ the frequency of an external electric field), we find that it scales as $\omega^{1-d_s/2}$ in the case of core segments. In order to have a reliable proof of this feature, we extend the iterative methods of Ref. \cite{dolgushev15} to calculate the local autocorrelation functions, that allow us to consider very large structures. In our analysis we identify the most important relaxation times, here the VF symmetry enables us to provide a well-performing approximate expression for these times.
 
The paper is structured as follows: In Sec.~\ref{model} we present the HP model and the corresponding local dynamical quantities, while the iterative methods for their computation are relegated to Appendix~\ref{AppA}. In Sec.~\ref{results} we provide and discuss our results. The paper ends with conclusions (Sec.~\ref{conclusions}). 

\section{Theory}\label{model}

\subsection{The structure of regular hyperbranched polymers}

We study hyperbranched polymers modeled by deterministic fractal structures. There are many fractal generators, see, e.g., Refs. \cite{vicsek92,feder13}. Among them, Vicsek fractals (VF) are of special interest, because their parametrization allows one to change intrinsic fractal properties, such as the spectral and fractal dimensions as well as to advance analytic calculations \cite{blumen03,blumen04}.

The topological structure of a VF is characterized by two parameters: the generation number $G$ and the functionality of the branching nodes, $F$. At generation $G$, a VF consists of $N=(F+1)^G$ beads. 
The density of states is of non-Debye kind, i.e., it scales as $\rho(\lambda)\sim\lambda^{\frac{d_s}{2}-1}$ with the spectral dimension $d_s$ \cite{alexander82}, which for VF is given by \cite{blumen03,blumen04}
\begin{equation}\label{d_s}
d_s=\frac{2\ln(F+1)}{\ln(3F+3)}.
\end{equation}

\begin{figure}[!ht]
\begin{center}
\includegraphics[width=8.5cm] {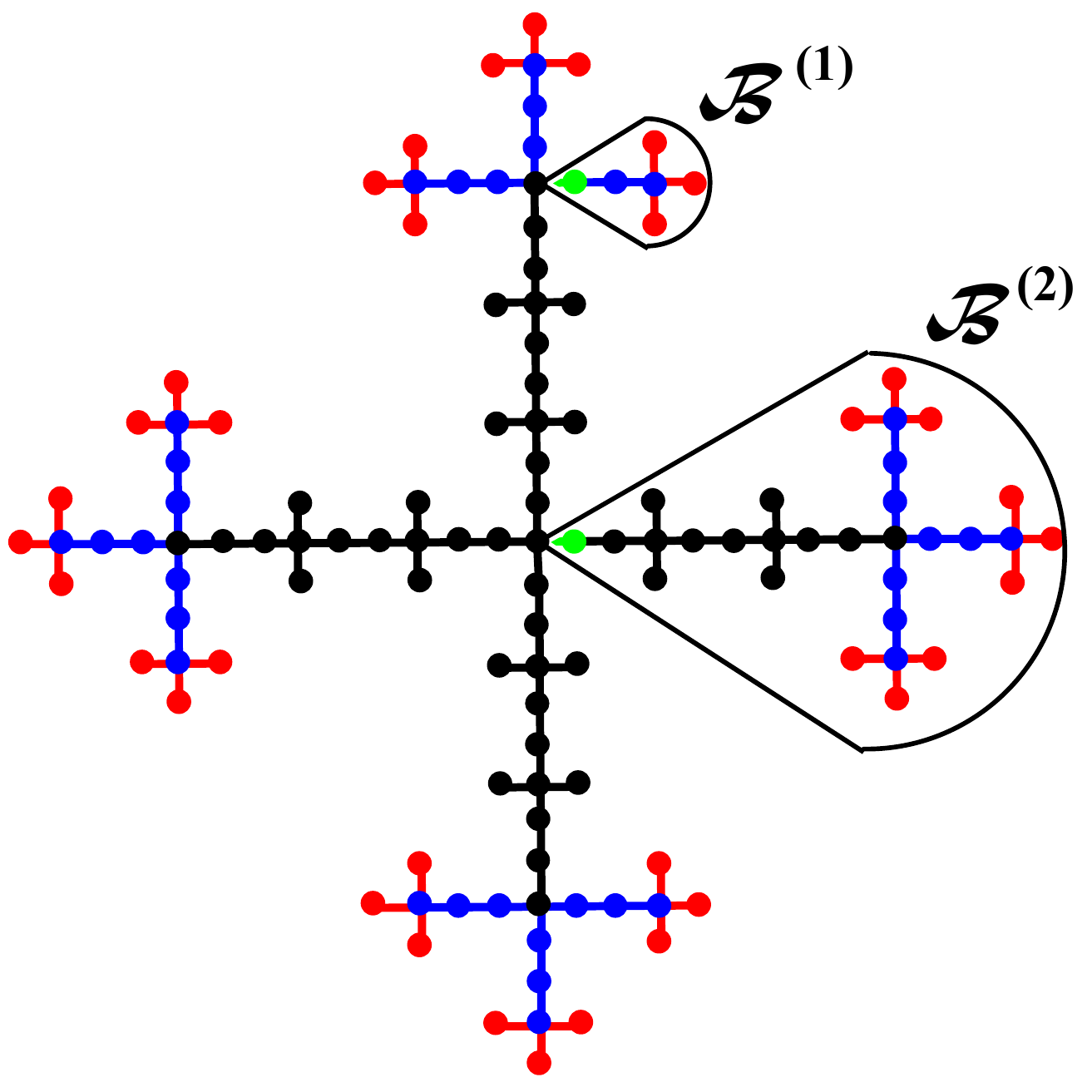}
\caption{(color online) Structure of VF of $F=4$ and generation $G=3$. Different shells $m$ are color-coded: $m=0$ (red), $m=1$ (blue), and $m=2$ (black). The drop-like shapes indicate examples of branches $\mathcal{B}^{(i)}$. Their root segments are color-coded by green.} \label{fig1}
\end{center}
\end{figure}

In this paper, we study the dynamics of bonds located at various positions in the fractal macromolecule. It is then useful to divide the structure into different shells, see Fig.~\ref{fig1}. We introduce the structural parameter $m$ which numbers different shells; the peripheral shell is associated with $m=0$, so that the shell related to the core is numbered by $m=G-1$. 

We define as a root segment of the shell $m$ any bond linking one of the most interior monomers of the shell $m$ to the most exterior monomer of the shell $m+1$. A root segment associated to $m=0$ is therefore a peripheral bond, while a root segment of the $(G-1)$th shell is one of the $F$ core segments. In the following, we will study the dynamics of these root segments. We also introduce the substructures originated from root segments, which we call "branch" and denote by $\mathcal{B}^{(i)}$, see Fig.~\ref{fig1}.

\subsection{Dynamical model}

The VF structure is represented by $N$ beads connected by springs. The position of the $i$th bead at time $t$ is represented by a vector in 3D space, $\ve[r]_i(t)$, whose dynamics follows from the Langevin equation \cite{gurtovenko05}:
\begin{equation}\label{langevin}
 	\zeta \frac{\partial}{\partial t}\ve[r]_{i}(t)+K \sum_{j=1}^{N}A_{ij}\ve[r]_{j}(t)=\ve[f]_{i}(t),
\end{equation}
where $\mathbf{A}=(A_{ij})$ is the connectivity matrix that reflects the VF topology: $A_{ii}$ is equal to the functionality of bead $i$ (i.e., the number of beads directly attached to $i$), if beads $i$ and $j$ are connected $A_{ij}$ is equal to $-1${,} and to $0$ otherwise. Moreover, in Eq.~(\ref{langevin}), $\zeta$ is the friction coefficient, $K$ is the spring constant, and the stochastic forces $\{\mathbf{f}_i(t)\}$ follow the white noise statistics, i.e.,
\begin{align}\label{forces}
\langle f_{i\alpha}(t)f_{j\beta}(t')\rangle=2\ k_{B}T\ \zeta\ \delta(t-t')\delta_{ij}\delta_{\alpha\beta},
\end{align}
with $\alpha$ and $\beta$ Cartesian coordinates $x,y,z$.

The pathway to the solution of Eq.~(\ref{langevin}) lies in the diagonalization of the matrix $\mathbf{A}$. We will denote by $u_i^{(\lambda,n)}$ the $i$th coordinate of the $n$th (normalized) eigenvector $|\mathbf{u}^{(\lambda,n)})$ associated with the eigenvalue $\lambda$, whose degeneracy is $D_{\lambda}$. Based on the eigenvectors, the bead coordinates can be decomposed,
\begin{align}
\ve[r]_i(t)=\sum_{\lambda}\sum_{n=1}^{D_{\lambda}}u_i^{(\lambda,n)}\ \tilde{\ve[a]}_{\lambda,n}(t), \label{0592}
\end{align}
where the sum runs over all distinct eigenvalues $\{\lambda_i\}$.
The vectors $|\mathbf{u}^{(\lambda,n)})$ are orthonormal, hence the eigenmode amplitudes are given by 
\begin{align} 
\langle \tilde{a}_{\lambda,n,\alpha}(t)\tilde{a}_{\lambda',n',\beta}(t')\rangle=\frac{ k_BT}{ \lambda K} \delta_{\alpha,\beta}\delta_{n,n'}\delta_{\lambda,\lambda'}\ e^{-\lambda\vert t-t'\vert/\tau_0}, \label{CorrAlambdaq} 
\end{align}
where we have introduced the monomeric relaxation time $\tau_0=\zeta/K$.

\subsection{Orientational relaxation functions}

In this work we are interested in the local relaxation properties related to the segments (springs) connecting nearest-neighboring beads. If the segment $\mathbf{d}_a$ connects the nearest neighboring beads $q_1$ and $q_2$, then one has
\begin{align}
	\ve[d]_a(t)\equiv \ve[r]_{q_1}(t)-\ve[r]_{q_2}(t)\equiv \sum_{i=1}^N (\mathbf{G}^T)_{ai} \ve[r]_i(t) \label{defXandh},
\end{align} 
where $\mathbf{G}$ is the so-called incidence matrix \cite{biggs93}.

Inserting Eq.~(\ref{0592}) into Eq.~(\ref{defXandh}) and using Eq.~(\ref{CorrAlambdaq}) we obtain the temporal autocorrelation function \cite{perico85,markelov14}
\begin{equation}\label{M1}
 M_1^{a}(t)\equiv\langle\mathbf{d}_a(t)\cdot\mathbf{d}_a(0)\rangle/l^2=\sum_{\lambda}C_{\lambda}^a\exp[-t/\tau_{\lambda}],
\end{equation}
where the sum over $\lambda$ runs over the $3^G-1$ distinct nonvanishing eigenvalues, $l^2=3k_B T/K$ is the mean-squared bond length, and $\tau_{\lambda}\equiv\tau_0/\lambda$. Moreover,  $C^a_{\lambda}$ is given by
\begin{equation}\label{Ck}
C_{\lambda}^a=\sum_{n=1}^{D_{\lambda}}\left[((\mathbf{G})_a|\mathbf{u}^{(\lambda,n)})\right]^2/\lambda.
\end{equation}
Here $((\mathbf{G})_a|\mathbf{u}^{(\lambda,n)})$ denotes the scalar product of the $a$th column of matrix $\mathbf{G}$ with the vector $|\mathbf{u}^{(\lambda,n)})$. 

In practice, in the case of large structures, the calculation of $C_{\lambda}^a$ is almost impossible by using brute force diagonalization of the matrix $\mathbf{A}$. Fortunately, symmetric fractal structures such as VF can be constructed iteratively from one generation to the next one, a fact that can be exploited to compute iteratively the eigenvalues \cite{blumen04}. It turns out that the coefficients $C_\lambda^a$ can also be computed iteratively by adapting  the projection operator techniques that were proposed in Ref. \cite{dolgushev15}. The description of this iterative method is rather technical and is left to the Appendix  \ref{AppA}, where we show how to extend the method proposed in Ref.  \cite{dolgushev15} to actual computation of the coefficients $C_\lambda^a$ for large VF macromolecules. 

For type A polymers in Stockmayer's classification~\cite{stockmayer67}, in which the dipole moments are aligned along polymers' segments \cite{Note1}, the $M_1^a$ function is closely related through the Fourier-Laplace transform to the frequency-dependent complex dielectric susceptibility $\Delta\epsilon^{*}_a(\omega)=(\epsilon^{*}(\omega)-\epsilon_{\infty})/(\epsilon_{0}-\epsilon_{\infty})$, with $\epsilon_{0}$ and $\epsilon_{\infty}$  the limiting low- and high-frequency dielectric constants, respectively.
One gets \cite{glarum60,perico85}  
\begin{equation}\label{de_gen}
 \Delta\epsilon^{*}_a(\omega)=-\int_0^{\infty}dt\frac{\mathrm{d}P_1^a(t)}{\mathrm{d}t}e^{-\mathrm{i}\omega t}\approx-\int_0^{\infty}dt\frac{\mathrm{d}M_1^a(t)}{\mathrm{d}t}e^{-\mathrm{i}\omega t}.
\end{equation}
Here the function $P_1^a(t)$ is the first Legendre polynomial $P_1^a(t)\equiv\langle(\mathbf{d}_a(t)\cdot\mathbf{d}_a(0))/(|\mathbf{d}_a(t)||\mathbf{d}_a(0)|)\rangle$ which is conventionally approximated by a well-working relation $P_1^a(t)\approx M_1^a(t)$.  The real and imaginary part of $\Delta\epsilon_a^{*}(\omega)=\Delta\epsilon'_a(\omega)-\mathrm{i}\Delta\epsilon''_a(\omega)$ are
\begin{equation}\label{diel_stor}
\Delta\epsilon'_a(\omega)=\sum_{\lambda}\frac{C_{\lambda}^a}{1+(\omega\tau_{\lambda})^2}
\end{equation}
and
\begin{equation}\label{diel_loss}
\Delta\epsilon''_a(\omega)=\sum_{\lambda}\frac{C_{\lambda}^a\omega\tau_{\lambda}}{1+(\omega\tau_{\lambda})^2}.
\end{equation}
In Eqs.~(\ref{diel_stor})-(\ref{diel_loss}) the sums run over distinct nonvanishing eigenvalues. Since we will compare the local and structure-averaged dynamics, we also consider the macroscopic dielectric relaxation, which is independent of the eigenvectors,
\begin{equation}\label{diel_stor_full}
\Delta\epsilon'(\omega)\equiv\frac{1}{N}\sum_a\Delta\epsilon'_a(\omega)=\frac{1}{N}\sum_{\lambda}\frac{D_{\lambda}}{1+(\omega\tau_{\lambda})^2}
\end{equation}
and
\begin{equation}\label{diel_loss_full}
\Delta\epsilon''(\omega)\equiv\frac{1}{N}\sum_a\Delta\epsilon''_a(\omega)=\frac{1}{N}\sum_{\lambda}\frac{D_{\lambda}\omega\tau_{\lambda}}{1+(\omega\tau_{\lambda})^2},
\end{equation}
see, e.g., Eqs.~(41) and (42) of Ref.~\cite{gurtovenko05}. In Eqs.~(\ref{diel_stor_full})-(\ref{diel_loss_full}) $D_{\lambda}$ denotes the degeneracy of the eigenvalue $\lambda$ and the sums run over distinct nonvanishing eigenvalues.

\section{Results and Discussion}\label{results}

\subsection{Relaxation of segments}

\begin{figure}[!ht]
\begin{center}
\includegraphics[width=8.5cm] {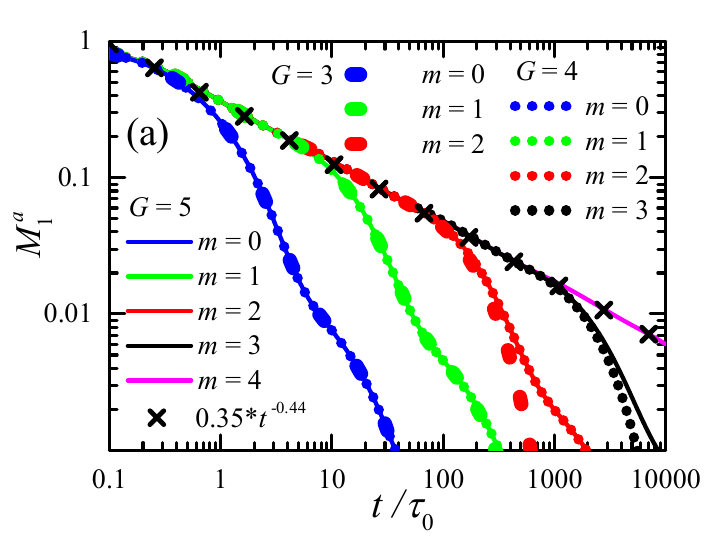}
\includegraphics[width=8.5cm] {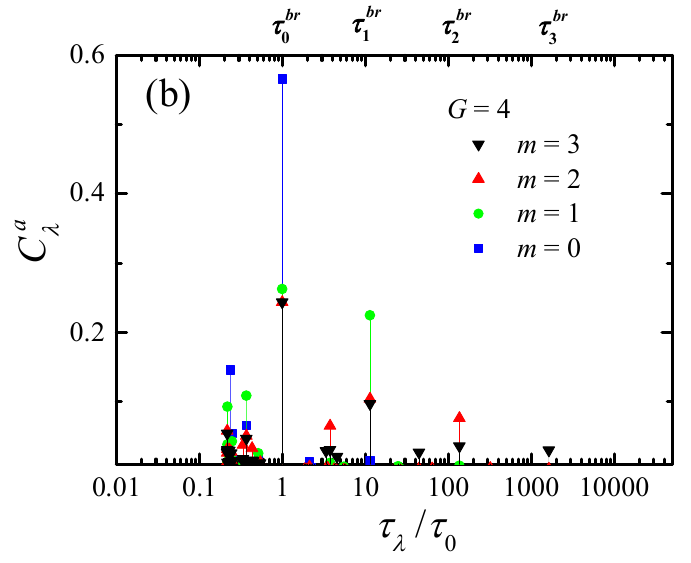}
\caption{(color online) (a) $M_1^a(t)$ functions for the root segments of different VF shells $m$. The structures are of different size $G$ and of $F=3$. (b) The contributions (Eq.~(\ref{Ck})) of the relaxation decays in $M_1^a(t)$ of (a) for all relaxation times $\tau_\lambda=\tau_0\lambda^{-1}$.}\label{fig2}
\end{center}
\end{figure}

We first describe the segment autocorrelation function $M_1^a(t)$ of Eq.~(\ref{M1}), focusing on the root segments of different shells. We find that the $M_1^a(t)$ function depends on the remoteness of the bond $\mathbf{d}_a$ from the periphery of the VF, i.e., on the parameter $m$; see Fig.~\ref{fig2}(a). 
As can be inferred from the figure, for the same value of $m$ the curves $M_1^a(t)$ overlap each other (apart from minor differences for $m=G-1$ at long times) while they correspond to different structure sizes (i.e., different generation $G$). For short times the decay of the $M_1$ is the same for all curves. For longer times, we observe a different picture: the decay of $M_1$ is faster for segments belonging to more peripheral shells. It can be also observed in Fig.~\ref{fig2}(a), especially for large $m$ and $G$, that the curves scale, $M_1^a(t)\sim t^{\frac{d_s}{2}-1}$, where $d_s$ is given by Eq.~(\ref{d_s}). The reason for this scaling is discussed later in Sec.~\ref{results}B.

\begin{figure}[!ht]
\begin{center}
\includegraphics[width=8.5cm] {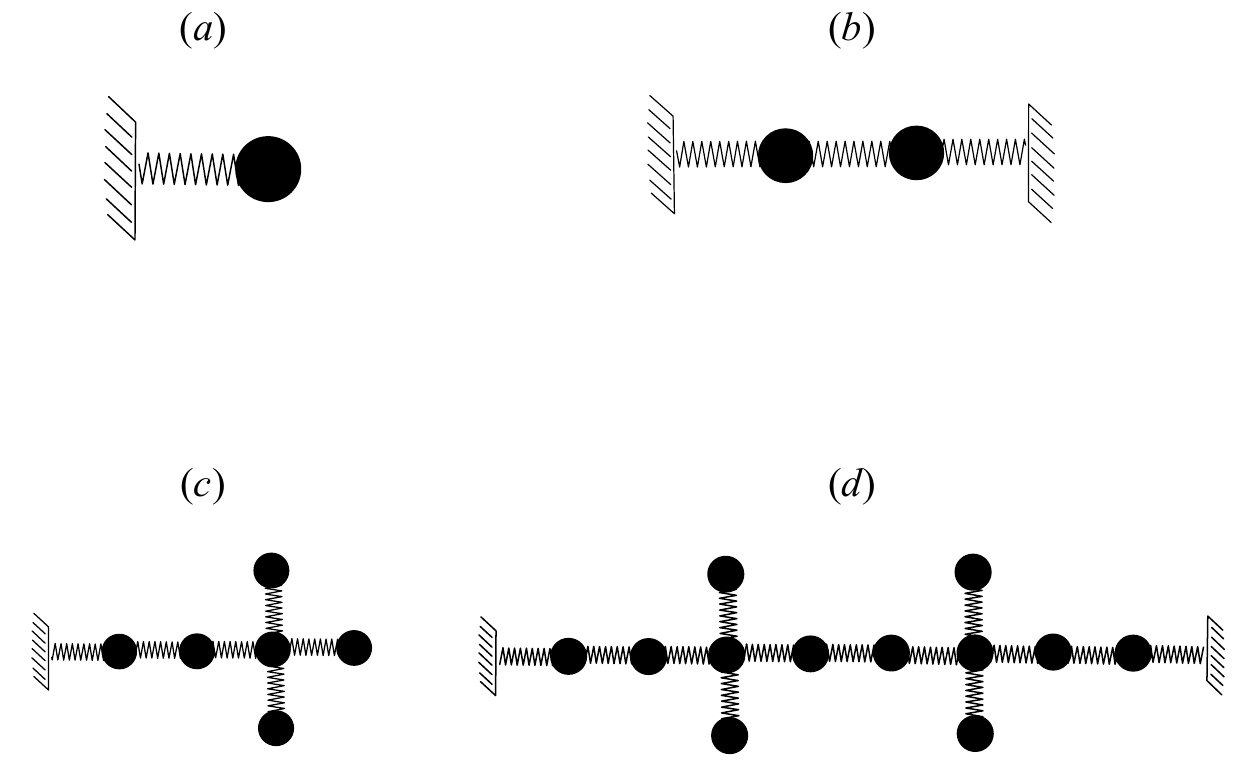}
\caption{The relaxation times of the symmetric modes of (b) and (d) coincide with the relaxation times of sets (a) and (c), respectively.}\label{fig3}
\end{center}
\end{figure}

In order to understand the behavior of of the $M_1$-functions, it is convenient to look at the contributions of different modes, see Fig.~\ref{fig2}(b) for $G=4$ and $F=3$. In this figure we see that there is a small number of relaxation times $\tau_\lambda$ for which the coefficients $C_\lambda^a$ admit large values and thus have significant contributions to the correlation function. It turns out that these modes are closely related to the global relaxation of different VF branches.   
We denote by $\tau_m^{br}$ the maximal relaxation time of a branch $\mathcal{B}^{(m)}$. As has been shown in Ref.~\cite{fuerstenberg13} $\tau_m^{br}$ is related to the mode in which two directly connected branches of type $\mathcal{B}^{(m)}$ move against each other as a whole (while the others are immobile). 
We remark on Fig.~\ref{fig2}(b) that the modes related to the time $\tau_m^{br}$ are not excited for segments from the shell $m'>m$. On the other hand, the dynamics of a segment of the shell $m$ significantly depends on the contributions corresponding to smaller branches, associated to the times $\tau_0^{br},\dots,\tau_m^{br}$. The reason for this behavior can be understood by looking on the examples illustrated on Fig.~\ref{fig3}: 
The relaxation time $\tau_0^{br}=\tau_0$ is related to the eigenmode in which two next nearest neighbor beads move with the same amplitude but against each other, while all other beads remain immobile \cite{fuerstenberg13}. Thus, it leads to a single degree of freedom \cite{fuerstenberg13} and could be visualized through Fig.~\ref{fig3}(a). However, as has been discussed in Ref. \cite{blumen04}, the relaxation time $\tau_0^{br}$ is also responsible for other modes, as soon as all beads of functionality $F$ remain immobile. Such a localization of modes is a fundamental feature for fractal systems \cite{yakubo89,devries89,bunde92,blumen04,granek05,reuveni12pre,zemlyanov92,zhang98}. In fact, $\tau_0^{br}$ is also one of four relaxation times of $\mathcal{B}^{(1)}$ for the case of immobile bead of functionality $F$. Looking precisely at this mode one finds that the beads between two immobile branching nodes move with the same amplitude and direction. So, 
the relaxation time $\tau_0^{br}$ related to Fig.~\ref{fig3}(a) is also a solution (related to the symmetric mode) for the system of Fig.~\ref{fig3}(b). Analogously, the maximal relaxation time for systems of Fig.~\ref{fig3}(c) and Fig.~\ref{fig3}(d) is $\tau_1^{br}$. In this way for a segment from the $m$th shell the times $\tau_0^{br},\dots,\tau_m^{br}$ turn out to be very important (see Fig.~\ref{fig2}(b)).

In the following we focus on the analysis of the relaxation times $\tau_m^{br}$. As has been shown in Refs.~\cite{blumen03,blumen04}, the eigenvalue spectrum of the connectivity matrix $\mathbf{A}$ and hence the relaxation times can be found for VF in an iterative way, see Eq.~(\ref{CubicEquation}) of Appendix~\ref{AppA}.  The $x=\tau_{k+1}^{br}/\tau_0$ is the maximal root of the polynomial equation involving $\tau_{k}^{br}/\tau_0$:
\begin{equation}\label{eq_pol}
(1-3x)(1-(F+1)x)=\frac{\tau_0}{\tau_{k}^{br}}x^3, 
\end{equation}
and the iteration is initialized with $\tau_0^{br}=\tau_0$. 
The above cubic polynomial equation can be readily solved. However, the resulting Cardano solutions \cite{blumen03,blumen04} are quite bulky. Here we provide a simple and accurate approximate expression for $\tau_k^{br}$, which follows by considering small $1/x$ expansions. More precisely, we approximate $\tau_{1}^{br}/\tau_0$ by the maximal solution of (\ref{eq_pol}) after expansion of the left hand side at second order in $1/x$. For all larger values $k\ge2$, we approximate the left hand side of (\ref{eq_pol}) at leading order in $1/x$, bringing to $\tau_{k+1}^{br}\simeq3(F+1)\tau_k^{br}$. 
The resulting approximation for $\tau_{k}^{br}$ reads:
\begin{equation}\label{tau_approx}
\tau_k^{br}\approx\frac{2(F+4)[3(F+1)]^{k-1}}{3(F+1)-\sqrt{9F^2+14F-7}}\tau_0. 
\end{equation}
which holds for $G-1\geq k>0$. In Appendix~\ref{AppB} we show that the accuracy of the expression (\ref{tau_approx}) is better than $0.5\%$. 

We remark, that the branch $\mathcal{B}^{(k)}$ consists of ${n=[(F+1)^{k+1}-1]/F}$ beads; hence the time $\tau_k^{br}$ grows faster than that for dendrimers (for which $\tau^{br} \sim n$) \cite{gotlib07,markelov14} and slower than for linear chains (for which $\tau_{max} \sim n^2$) \cite{doi88}.

\begin{figure}[!ht]
\begin{center}
\includegraphics[width=8.5cm] {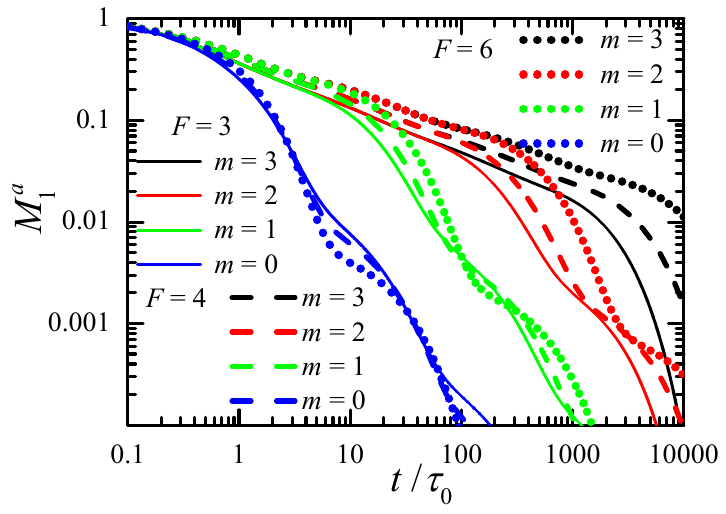}
\caption{(color online) The $M_1(t)$ functions for VF of generation $G=4$ and different $F$. The results are presented for root segments belonging to different shells indicated by $m$.}\label{fig4}
\end{center}
\end{figure}

As it follows from Eq.~(\ref{tau_approx}), the functionality $F$ plays an important role for the branch relaxation. Therefore the
 role of branch size is reflected in the $M_1^a(t)$-functions, see Fig.~\ref{fig4}. For higher $F$ the corresponding branch $\mathcal{B}^{(m)}$ is larger, leading to larger $\tau_m^{br}$. Therefore for VF of higher $F$ the $M_1^a(t)$ show a slower decay. 

We note that the degeneracy $D_k$ of $\tau_k^{br}$ reads \cite{blumen04}
\begin{equation}\label{degeneracy}
D_k=(F-2)(F+1)^{G-k-1}+1.
\end{equation}

We close this subsection by remarking that such an analysis can be also done for side substructures. Also in this case the size of the side branches plays a key role.

\subsection{Dielectric relaxation and scaling}

\begin{figure}[!ht]
\begin{center}
\includegraphics[width=8cm] {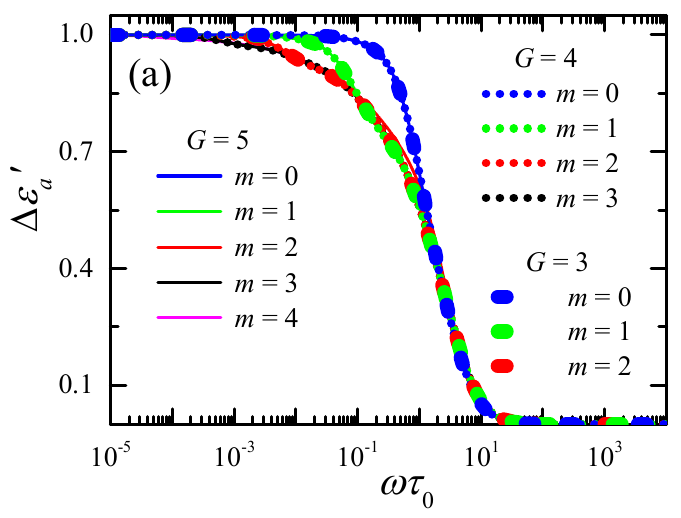}
\includegraphics[width=8.5cm] {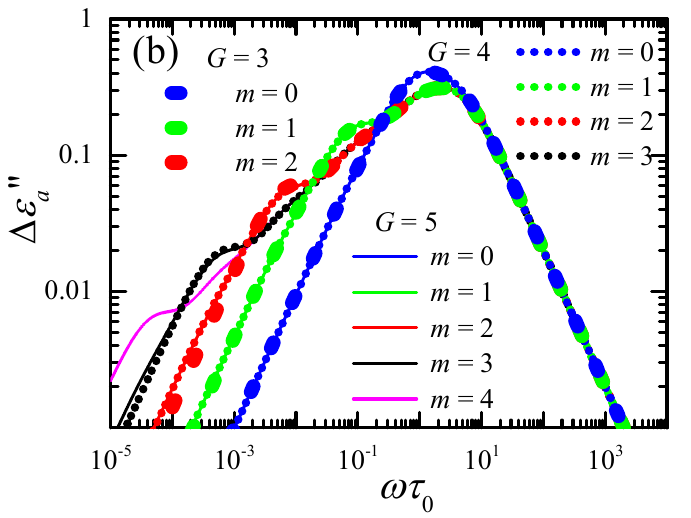}
\caption{(color online) The real part $\Delta\epsilon'_a(\omega)$  and the imaginary part $\Delta\epsilon''_a(\omega)$ of the complex dielectric susceptibility for single segments {of VF of different size $G$ and of $F=3$. The choice} of the segments is the same as in Fig.~\ref{fig2}.}\label{fig5}
\end{center}
\end{figure}

The segment autocorrelation functions are closely related to the dielectric relaxation functions, see Eqs.~(\ref{diel_stor})-(\ref{diel_loss}). In Fig.~\ref{fig5} we plot $\Delta\epsilon'_a(\omega)$ and $\Delta\epsilon''_a(\omega)$. We remark that, at high frequencies, the curves for different shells $m$ overlap each other. This means that the segment dynamics at short times is sensitive only to the local neighborhood, i.e., it is independent of the size of the VF branch. 
At smaller frequencies, the curves differ, especially in $\Delta\epsilon''_a(\omega)$. In this region, the dynamics reflects the branch relaxation times $\tau^{br}_k$. Hence for higher $m$ the $\Delta\epsilon''_a(\omega)$ function decays at lower frequencies. 
Also as for $M_1(t)$ there is practically no difference for different $G$ but the same $m$.

\begin{figure*}[!ht]
\begin{center}
\includegraphics[width=17cm] {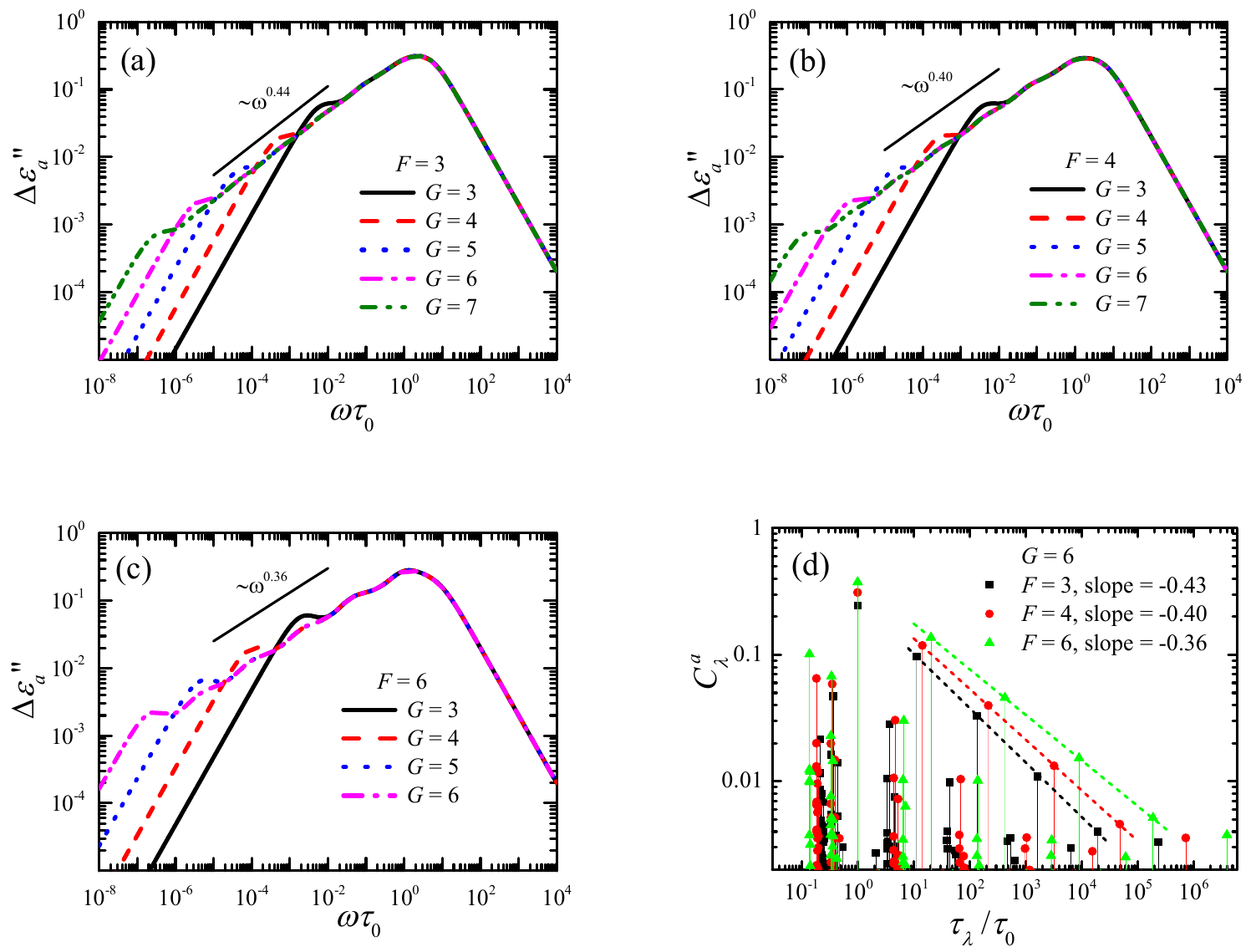}
\caption{(color online) Plots (a)-(c): The imaginary part $\Delta\epsilon''_a(\omega)$ of the complex dielectric susceptibility for the core segments of VF for different parameters $F$ and $G$. Plot (d): Contributions $C_\lambda^a$ for $\Delta\epsilon''_a(\omega)$ of plots (a)-(c), the size of VF is $G=6$.}\label{fig6}
\end{center}
\end{figure*}

Now, as can be observed in Fig.~\ref{fig5}, for higher $m$, a scaling behavior develops. Let us consider the core segments for different functionalities $F$, see Fig.~\ref{fig6}. Fitting the $\Delta\epsilon''_a(\omega)$ curves with power-laws in the intermediate frequency region yields for $F=3$, $4$, and $6$ the exponents around $0.44$, $0.40$, and $0.36$, respectively. These values are close to $1-\frac{d_s}{2}$ where $d_s$ is the spectral dimension of VF  given by Eq.~(\ref{d_s}).

To investigate the origin of this scaling, let us transform the discrete sum in Eq.~(\ref{diel_loss}) into an integral,
\begin{equation}\label{diel_loss_cont}
\sum_{\lambda}\frac{C_{\lambda}^a\omega\tau_{\lambda}}{1+(\omega\tau_{\lambda})^2}\Rightarrow\int_{\lambda_{\min}}^{\lambda_{\max}}\mathrm{d}\lambda\frac{\rho(\lambda)}{D(\lambda)}\frac{C_{\lambda}^a\omega\tau_{\lambda}}{1+(\omega\tau_{\lambda})^2},
\end{equation}
where the spectral density $\rho(\lambda)$ is normalized by the degeneracy $D(\lambda)$ related to the eigenvalue $\lambda$. According to the definition of spectral dimension $d_s$ the relation $\rho(\lambda)\sim\lambda^{\frac{d_s}{2}-1}$ holds \cite{alexander82}. The behavior of $D(\lambda)$ comes from inspection of Eqs.~(\ref{tau_approx})-(\ref{degeneracy}): Reminding that $\tau_{\lambda}=\tau_0/\lambda$, for small nonvanishing $\{\lambda_j\}$ numbered in ascending order follows $\lambda_j\sim(3F+3)^j/(3F+3)^G$ and $D(\lambda_j)\sim(F+1)^j$. Thus, $D(\lambda)\sim\lambda^{\frac{d_s}{2}}$ with $d_s$ from Eq.~(\ref{d_s}). Furthermore, the numerical analysis of $C_\lambda^a$ shows that $C_\lambda^a\sim\tau_{\lambda}^{\frac{d_s}{2}-1}\sim\lambda^{1-\frac{d_s}{2}}$, see Fig.~\ref{fig6}(d). The reason for this scaling lies in the behavior of the function $\left[((\mathbf{G})_a|\mathbf{u}^{(\lambda,n)})\right]^2$ of Eq.~(\ref{Ck}). It behaves as $\sim\lambda^{\frac{d_s}{d_l}}$, see, e.g., Eqs.~(10,22) of Ref. \cite{reuveni12pre} keeping in mind the relation between $\omega$ of Ref. \cite{reuveni12pre} and $\lambda$, $\lambda\sim\omega^2$. For fractals without non-Alexander-Orbach anomaly \cite{sokolov16} the topological dimension $d_l$ (which is for treelike structures equivalent to their fractal dimension in the stretched state) is related to the spectral dimension as $d_l=d_s/(2-d_s)$ \cite{alexander82,cates84,muthukumar85}. Thus, $\lambda C_\lambda^a/D(\lambda)\sim\lambda^{2-d_s}$, from which  $C_\lambda^a\sim\lambda^{1-\frac{d_s}{2}}$ follows. Therefore,
\begin{equation}\label{diel_loss_scal}
\Delta\epsilon''_a(\omega)\sim\int\frac{\mathrm{d}\lambda}{\lambda^{\frac{d_s}{2}}}\frac{\lambda/\omega}{1+(\lambda/\omega)^2}\sim\omega^{1-\frac{d_s}{2}},
\end{equation}
in line with the numerical observations. Note that the fact that the modes associated to $\tau_0^{br},\tau_1^{br},...$ dominate the dynamics is checked explicitly in Appendix~\ref{AppB}.

Performing similar calculations for the $M_1^a(t)$-functions of the same segments, we obtain
\begin{equation}\label{M1_scal}
M_1^a(t)\sim\int\mathrm{d}\lambda\frac{\rho(\lambda)}{D(\lambda)}C_{\lambda}^a\,e^{-\lambda t/\tau_0}\sim t^{\frac{d_s}{2}-1}.
\end{equation}
In Fig.~\ref{fig2}(a) we show that the scaling of Eq.~(\ref{M1_scal}) is in good agreement with numerical calculations, which is also consistent with the scaling analysis of Refs. \cite{reuveni12,reuveni12pre}.

\subsection{Comparison with different structure-averaged dielectric relaxations}
There is a special interest in the dielectric relaxation averaged over segments of the same type, i.e., segments connecting bead pairs of the same (pairwise) functionalities. Such segments will have different chemical structure so that they may be recognizable in experiments, as it was present, e.g., for Fr\'{e}chet dendrimers \cite{hecht99,pinto13a}. For VF, bonds can connect pairs of beads of functionalities $(F,1)$, $(F,2)$, or $(2,2)$, thereby defining 3 different types of bonds. 

\begin{figure}[!ht]
\begin{center}
\includegraphics[width=8.5cm] {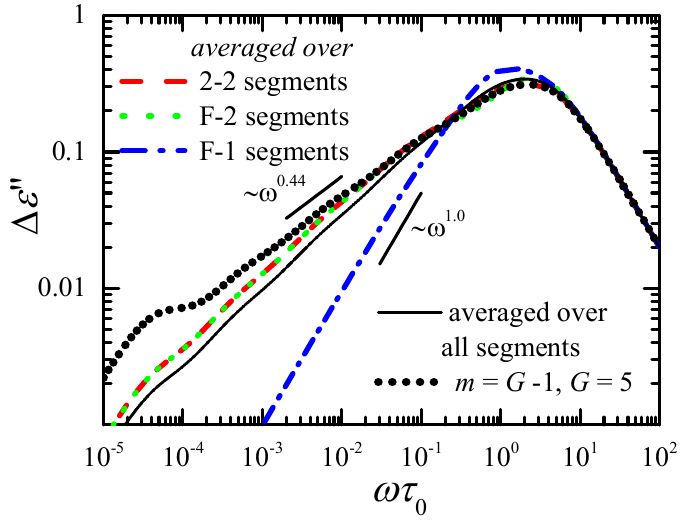}
\caption{(color online) Imaginary part of the dielectric susceptibility averaged over segments of the same type. The VF is of generation $G=5$ and $F=3$.}\label{fig7}
\end{center}
\end{figure}

In Fig.~\ref{fig7} we display the $\Delta\epsilon''(\omega)$ functions for all segments of the same kind, i.e. for segments of $(F_1,F_2)$-type we have
 \begin{equation}\label{diel_loss_segment}
\Delta\epsilon''_{(F_1,F_2)}(\omega)=\frac{1}{N_{(F_1,F_2)}}\sum_{a\in(F_1,F_2)}^{N_{(F_1,F_2)}}\Delta\epsilon''_a(\omega),
\end{equation}
where $N_{(F_1,F_2)}$ is the number of segments of type $(F_1,F_2)$. We also display on Fig.~\ref{fig7} the $\Delta\epsilon''(\omega)$ obtained after averaging over all segments [see Eq.~(\ref{diel_loss_full})].

The analysis of the intermediate frequency region on Fig.~\ref{fig7} shows that different type of segments lead to qualitatively different dynamics. The segments of $(F,1)$-kind show a very quick dynamics. The relaxation of these segments is characterized through a single relaxation time $\tau_0$ and hence the dielectric relaxation function has a trivial shape. On the contrary, the segments of other type show a rich behavior. The functions $\Delta\epsilon''(\omega)$ averaged over all segments of $(F,2)$ and of $(2,2)$ kind  follow closely the behavior of the root segment at high intermediate frequencies. However, for lower frequencies they tend to the macroscopic, overall behavior which carries a different scaling behavior. Indeed, as it was shown, e.g., in Ref.~\cite{gurtovenko02}, summing up over all bonds one gets $\sum_a\left[((\mathbf{G})_a|\mathbf{u}^{(\lambda,n)})\right]^2=\lambda$. With this relation, following the steps of Eq.~(\ref{diel_loss_scal}), one obtains for structurally averaged $\Delta\epsilon''(\omega)$ the scaling $\Delta\epsilon''(\omega)\sim\omega^{d_s/2}$, see Ref.~\cite{blumen04}.
This scaling reflects, as for the generalized Landau-Peierls instability \cite{burioni02,granek05,reuveni12pre}, the global behavior of vibrations and differs from the scaling observed in Fig.~\ref{fig6} for core segments, for which the slope is $1-\frac{d_s}{2}$. Thus, we can summarize that different types of segments clearly manifest themselves through different scaling laws, and that the local bond dynamics can be very different from structure-averaged dynamics.

\section{Conclusions}\label{conclusions}

In this work we have studied the local segment dynamics in hyperbranched polymers modeled by Vicsek fractals (VF), focusing on the autocorrelation functions and the dielectric relaxation forms of single segments. We have found that the dynamics of segments is strongly related to their location in the structure. For the root segments of different shells under investigation, we found that the size of the whole macromolecule is rather unimportant, since segments at the same distance from the periphery show similar dynamic behavior for structures of very different sizes.

By analyzing the contributions of different relaxation times, we found that for the root segments of the $m$th shell (numbered from the periphery) the times related to the relaxation of branches as whole play a significant role. Moreover, if a smaller time has the same value as the relaxation time of a smaller branch, it will contribute significantly to the local relaxation (in analogy to the interchain relaxation spectrum in networks \cite{gurtovenko98} and different from dendrimers \cite{gotlib07,markelov14}). For all these fundamental times we found a well-working approximate expression. 

Remarkably, the analysis of the scaling behavior revealed that the core segments show a dynamics that is slower than the overall dynamics. For the imaginary part of the complex dielectric susceptibility, the corresponding exponent is given by $1-\frac{d_s}{2}$ as compared with $d_s/2$ for the overall dynamics, where $d_s$ is the spectral dimension. Note that this difference of scaling between local and structure-averaged dynamics does not appear for chains (where $d_s=1$) and it is instead a characteristic feature of hyperbranched macromolecules.

As a methodological point, we have provided here iterative methods for calculation of the bond autocorrelation functions. These methods, however, can be transferred to computation of other dynamic properties involving, for instance, not only the eigenvalues but also the eigenvectors of the dynamical matrix. Such problems naturally appear when one is interested in the dynamics of some part of the system rather than in the averaged, macroscopic overall evolution. In particular, for structures with local heterogeneity causing violation of the Alexander-Orbach relation \cite{sokolov16} such aspects are of great importance.

\acknowledgments

The authors thank Alexander Blumen {and Vladimir Toshchevikov} for fruitful discussions. M.D. acknowledges the support through Grant No. GRK 1642/1 of the Deutsche Forschungsgemeinschaft. D.A.M. acknowledges the Russian Foundation for Basic Research (grant No. 14-03-00926) and the Government of the Russian Federation (grant 074-U01). 

\appendix
\section{Reaching large structure sizes: iterative method for the computation of the bond correlation functions}
\label{AppA}

The computation of the correlation function requires the knowledge of the eigenvectors of the dynamical matrix $\mathbf{A}$ [see Eq.~(\ref{Ck})]. However, for large structures, this cannot be achieved from brute force diagonalization. Here we extend the iterative procedure of Ref. \cite{dolgushev15} that enables the computation of the amplitudes $C_{\lambda}^a$.

We start by rewriting Eq.~(\ref{Ck}) as
\begin{equation}\label{lambdaC}
\lambda C_{\lambda}^a=\sum_{n=1}^{D_{\lambda}}\left[((\mathbf{G})_a|\mathbf{u}^{(\lambda,n)})\right]^2=((\mathbf{G})_a|\hat{P}_{\lambda}|(\mathbf{G})_a), 
\end{equation}
where we have introduced the projection operator $\hat{P}_{\lambda}$,
\begin{equation}\label{def_P}
\hat{P}_{\lambda}=\sum_{n=1}^{D_{\lambda}} |\ve[u]^{(\lambda,n)})(\ve[u]^{(\lambda,n)}|.
\end{equation}
Consider now a list   $\{|\ve[w]^{(\lambda,q')})\}$  of $N-D_{\lambda}$ linearly independent (but not necessarily orthogonal) vectors that form a basis of the  eigensubspace associated with $\lambda$, i.e., that satisfy $(\ve[w]^{(\lambda,q')}| \ve[u]^{(\lambda,q)})=0$ for all $q,q'$. Define now a $(N-D_{\lambda})\times N$  matrix $\mathbf{W}_{\lambda}$ constructed from these vectors as  $\mathbf{W}_{\lambda}\equiv(|\ve[w]^{(\lambda,(D_{\lambda}+1))}),\dots,|\ve[w]^{(\lambda,N)}))^T$, where $^T$ denotes transpose. 
With this matrix the complementary projection operator $\hat{Q}_{\lambda}\equiv\hat{I}-\hat{P}_{\lambda}$ reads \cite{meyer00}
\begin{equation}\label{def_Q}
\hat{Q}_{\lambda}=\mathbf{W}_{\lambda}^T(\mathbf{W}_{\lambda}\mathbf{W}_{\lambda}^T)^{-1}\mathbf{W}_{\lambda}.
\end{equation}
The fact that each column of $\mathbf{G}$ contains only two nonzero entries, namely, $+1$ and $-1$ \cite{biggs93}, implies that $((\mathbf{G})_a|(\mathbf{G})_a)=2$, hence we obtain
\begin{align}\label{b_lambda_final}
\lambda C_{\lambda}^a&=((\mathbf{G})_a|\hat{I}-\hat{Q}_{\lambda}|(\mathbf{G})_a)\nonumber \\
&=2-((\mathbf{G})_a|\mathbf{W}_{\lambda}^T(\mathbf{W}_{\lambda}\mathbf{W}_{\lambda}^T)^{-1}\mathbf{W}_{\lambda}|(\mathbf{G})_a).
\end{align}
Thus, in order to calculate $C_{\lambda}^a$ we need to construct the matrices $\mathbf{W}_{\lambda}$ involved in Eq.~(\ref{b_lambda_final}). For this we adapt a decimation procedure that was proposed in Ref. \cite{blumen04} for the computation of the eigenvalues $\lambda$.  

\begin{figure}[!ht]
\begin{center}
\includegraphics[width=8cm]{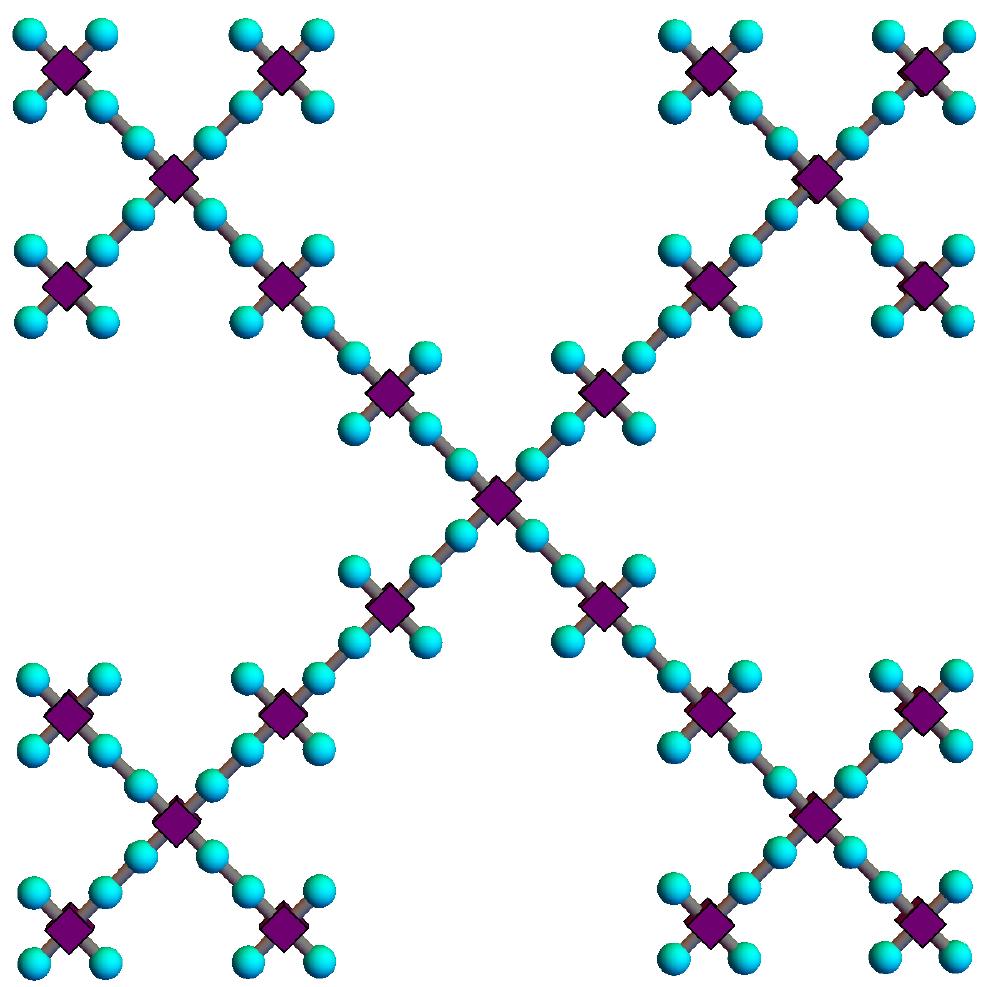}
\caption{(color online) Iterative construction of VF structures. Here, the structure with both the cubes and the spheres form a VF of generation $G=3$. In the decimated structure, where we keep only the branching nodes (cubes), ones obtains a VF of former generation $G=2$. All nodes with a spherical shape appeared at the  iteration from $G=2$ to $G=3$.} \label{figApp}
\end{center}
\end{figure}

The VF structure (see Fig.~\ref{figApp}) suggests that the next generation can be obtained from the previous one by attaching to each of the beads $F$ another beads. In the following we will denote the "old" beads (represented by cubes in Fig.~\ref{figApp}) by greek letters, say by $\mu$, and the "new" ones (represented by spheres in Fig.~\ref{figApp}) by latin letters, say by $k$.

\begin{table*}
\begin{center}
\begin{tabularx}{\textwidth}{c||ccc|ccc|ccc}
\hline\hline
$k$ & $F=3$, Eq.~(14) & $F=3$, Eq.~(15) & diff. & $F=4$, Eq.~(14) & $F=4$, Eq.~(15) &  diff. & $F=6$, Eq.~(14) & $F=6$, Eq.~(15) & diff.  \\\hline
$1$ & $11.393$ & $11.385$ &  $0.072\,\%$ & $14.451$ & $14.446$ &  $0.035\,\%$ & $20.515$ & $20.513$ & $0.012\,\%$ \\
$2$ & $136.1$ & $136.6$ &  $0.36\,\%$ &$216.2$ & $216.7$ &  $0.213\,\%$ & $430.3$ & $430.8$ & $0.1\,\%$\\
$3$ & $1633$ & $1639$ &  $0.40\,\%$ & $3243$ & $3250$ & $0.230\,\%$ & $9037$ & $9046$ & $0.1\,\%$ \\
$4$ & $19596$ & $19674$ &  $0.40\,\%$ &  $48644$ & $48756$ & $0.230\,\%$ & $189768$ & $189966$ & $0.1\,\%$\\ 
$5$ & $2.35\cdot10^{5}$ & $2.36\cdot10^{5}$ &  $0.40\,\%$ & $7.30\cdot10^{5}$ & $7.31\cdot10^{5}$ & $0.230\,\%$& $39.85\cdot10^{5}$ & $39.89\cdot10^{5}$ & $0.1\,\%$\\
$6$ & $2.82\cdot10^{6}$ & $2.83\cdot10^{6}$ &  $0.40\,\%$ & $10.94\cdot10^{6}$ & $10.97\cdot10^{6}$ & $0.230\,\%$ & $83.69\cdot10^{6}$ & $83.78\cdot10^{6}$ & $0.1\,\%$ \\
$7$ & $3.39\cdot10^{7}$ & $3.40\cdot10^{7}$ &  $0.40\,\%$ & $16.42\cdot10^{7}$ & $16.46\cdot10^{7}$ & $0.230\,\%$ & $175.7\cdot10^{7}$ & $175.9\cdot10^{7}$ & $0.1\,\%$ \\
$8$ & $4.06\cdot10^{8}$ & $4.08\cdot10^{8}$ &  $0.40\,\%$ & $24.63\cdot10^{8}$ & $24.68\cdot10^{8}$ & $0.230\,\%$ & $369.1\cdot10^{8}$ & $369.4\cdot10^{8}$ & $0.1\,\%$ \\
\hline\hline
\end{tabularx} 
\end{center}
\caption{Comparison of $\tau_k^{br}/\tau_0$ obtained from precise iterative calculations based on Eq.~(14) of the main text and from approximate Eq.~(15) of the main text.}\label{table} 
\end{table*}

Let $\ve[\Phi]^{(G)}=(\phi_1,...,\phi_N)$ be an eigenvector associated with the eigenvalue $\lambda^{(G)}$ of the matrix $\mathbf{A}$ of a VF of generation $G$. Now, collecting from $\ve[\Phi]^{(G)}$ only the entries related to the beads of previous generation (indicated by greek letters), one obtains for $\lambda^{(G)}\ne\{0,1,F+1\}$ an eigenvector $\ve[\Phi]^{(G-1)}=\{\phi_{\mu}\}$ of a VF of the previous generation $G-1$ \cite{blumen04}. Then the eigenvalue $\lambda^{(G-1)}$ associated with $\{\phi_{\mu}\}$ obeys the relation \cite{blumen04} 
\begin{align}
	\lambda^{(G-1)}=\lambda^{(G)}(\lambda^{(G)}-3)(\lambda^{(G)}-F-1)=P(\lambda^{(G)}).  \label{CubicEquation}
\end{align}
Given that the values of $\{\phi_{\mu}\}$ are already present at generation $G-1$, the values of $\phi_k$ for the "new" sites follow from the equations $\sum_{i=1}^N(A_{ki}-\lambda^{(G)}\delta_{ki})\Phi_i^{(G)}=0$. Hence, if one knows the matrix $\mathbf{W}_{\lambda^{(G-1)}}^{(G-1)}$, then the matrix $\mathbf{W}_{\lambda^{(G)}}^{(G)}$ at next generation is
\begin{align}
	\label{iterativeW}
\mathbf{W}^{(G)}_{\lambda^{(G)}}=
\begin{pmatrix}
\mathbf{W}^{(G-1)}_{\lambda^{(G-1)}} & \mathbf{0}\vspace{0.3cm}\\
\hspace{0.5cm}\mathbf{L}
\end{pmatrix},
\end{align}
where $\mathbf{L}$ stands for the lines of  $\mathbf{A}^{(G)}-\lambda^{(G)} \mathbf{I}$ related to the "new" sites $k$. With this relation one can readily construct the matrices $\ve[W]_{\lambda}$ iteratively. Since  Eq.~(\ref{CubicEquation})  is cubic,  the matrix $\mathbf{W}^{(G-1)}_{\lambda^{(G-1)}}$ will produce three matrices  $\ve[W]_{\lambda}$ at generation $G$, which can therefore be constructed iteratively.

The construction of matrices matrices $\ve[W]_{\lambda}$ is initialized by providing the matrix $\ve[W]_1^{(G)}$ associated with the eigenvalue $\lambda=1$ at generation $G$. For any eigenvector $\ve[\Phi]=(\phi_1,...,\phi_N)$ associated with the eigenvalue $\lambda=1$ holds $(\mathbf{A}-\mathbf{I})\bf{\Phi}=0$. From this, the equations for all the beads $\mu$ of functionality $F$ follow
\begin{align}
\phi_{\mu}=0,\label{eig1_f}
\hspace{0.4cm}\sum_{i\in \text{ NN of } \mu}\phi_{i}=0,
\end{align}
where the sum over $i$ runs over nearest-neighbors of site $\mu$. Furthermore, each pair of sites of functionality $2$ obeys the $(F+1)^{G-1}-1$ equations
\begin{equation}\label{eig1_2}
	\phi_{i}-\phi_{j}=0.
\end{equation}
The linearly independent Eqs. (\ref{eig1_f}) and (\ref{eig1_2}) determine the eigensubspace related to $\lambda=1$. Writing down these equations in the matrix form, $\ve[W]_1^{(G)}\ve[\Phi]=\ve[0]$, leads to matrix $\ve[W]_1^{(G)}$. 

Here, contrarily to Ref. \cite{dolgushev15}, the nondegenrate eigenvalues need to be considered explicitly. In the case of a non-degenerate eigenvalue $\lambda$, it is appropriate to compute the coefficients $C_{\lambda}^a$ with Eq.~ (\ref{lambdaC}), where the sum runs over only one term involving the eigenvector $\ve[w]_\lambda^G$ at generation $G$. This eigenvector can be computed by iteration. We note that, if $\ve[w]_i$ is an eigenvector at generation $G$, then it also implies that the vector $\ve[w]_\mu$, restricted to the sites $\mu$ of the decimated structure, is an eigenvector of the VF at generation $G-1$ associated with the eigenvalue $P(\lambda)$. Furthermore, using Eqs.~(56,57,67,68,69,76) of Ref. \cite{blumen04}, the coordinates $w_i$ of the new sites are given by
\begin{align}
w_i=w_\mu/(1-\lambda)  \label{ItW_i0}
\end{align}   
when $i$ is a terminal site of the new structure that is neighboring $\mu$ and 
\begin{align}
\label{ItW_i}
w_i=\frac{1}{(2-\lambda)^2-1}[(2-\lambda)w_\mu+w_{\mu'} ]
\end{align}   
otherwise. In the above equation,  $\mu'$ represent the site at former generation that is next nearest neighboring $i$ (see Fig.~\ref{figApp}).  

Therefore, using these formulas (\ref{ItW_i0},\ref{ItW_i}), all the eigenvectors associated to a nondegenerate eigenvalue at generation $G$ can be deduced from that at generation $G-1$, suggesting an iterative procedure. The procedure starts with the eigenvector associated to $\lambda=F+1$, for which one can readily show that $w_\mu=1$ (for all the sites of the decimated structure) and $w_i=-1/F$ for all other sites. 

\section{Accuracy of Eq.~(15) of the main text}
\label{AppB}

Here we look at the accuracy of the approximate Eq.~(15) of the main text. In Table~\ref{table} we present the values of the relaxation times $\tau_k^{br}$ calculated based on Eq.~(15) of the main text for $k=1\dots8$ and $F=3,4,6$ and compare them with the corresponding values obtained from the numerical solution of the iterative Eq.~(14) of the main text. The inspection of Table~\ref{table} shows that the maximal error appears for lower $F$ and it is not bigger than $0.5\%$.

\begin{figure*}[!ht]
\begin{center}
\includegraphics[width=17cm] {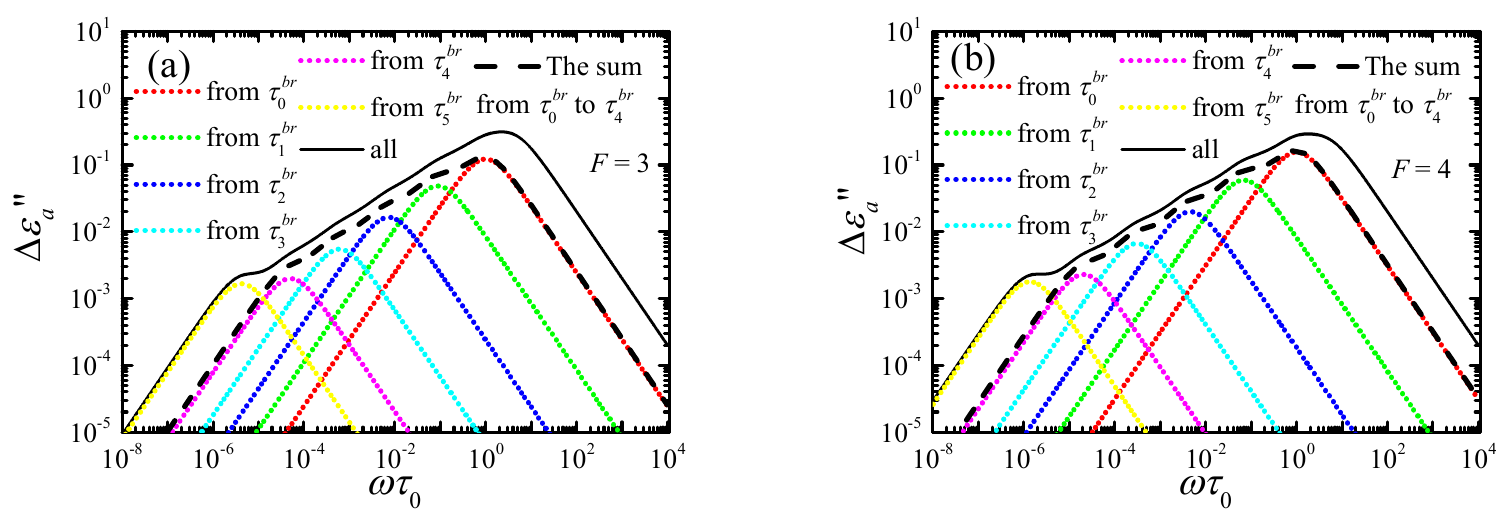}
\caption{(color online) The imaginary part $\Delta\epsilon''_a(\omega)$ and the contributions to it coming from the relaxation times $\{\tau_k^{br}\}$ of the complex dielectric susceptibility for the core segments of VF for $F=3$ and $4$ and $G=6$.}\label{fig9}
\end{center}
\end{figure*}

Let us check explicitly the role of the relaxation times  $\{\tau_k^{br}\}$ for the imaginary part of the complex dielectric susceptibility $\Delta\epsilon''_a(\omega)$  of VF core segments. In Fig.~\ref{fig9} we plot based on the $C_{\lambda}^a$ corresponding to the relaxation times $\{\tau_k^{br}\}$ the single curves for each of the relaxation times. As can be inferred from the figure, the summation of these contributions lead to a plot which is quite close to that obtained based on all  $\{\tau_{\lambda},C_{\lambda}\}$ and with the same scaling in the intermediate region of frequencies. This shows that the branch relaxation times $\{\tau_k^{br}\}$ dominate the dynamics of core segments.


%

\end{document}